\documentclass{article}

\usepackage{arxiv}

\usepackage[utf8]{inputenc}
\usepackage[english]{babel}
\usepackage[T1]{fontenc}
\usepackage{hyperref}
\usepackage{url}
\usepackage{booktabs}
\usepackage{amsfonts}
\usepackage{amsmath}
\usepackage{amssymb}
\usepackage{nicefrac}
\usepackage{microtype}
\usepackage{graphicx}
\usepackage{textcomp}
\usepackage{natbib}
\usepackage{doi}

\graphicspath{{figures/}}

\title{Astronomical Methods and Instrumentation in the Islamic World: Past, Present, Future}

\author{
  Armin Maghami~Asl \\
  Horace Mann School \\
  New York, NY \\
  \texttt{arminmasl26@gmail.com}
  \and
  Yaseen Almleaky \\
  Astronomy \& Space Sciences Dept. \\
  King Abdulaziz University, Jeddah \\
  \texttt{ymleaky2010@gmail.com}
}

\renewcommand{\shorttitle}{Astronomy in the Islamic World}

\hypersetup{
  pdftitle={Astronomy in the Islamic World},
  pdfauthor={Armin M. Asl, Yaseen Almleaky},
  pdfkeywords={Islamic astronomy, al-Sufi, al-Biruni, al-Battani, Ibn Sina, astrolabe, sextant, Maragha, Samarkand, supernovae, comets, eclipses, Andromeda, knowledge transmission, global history of science, astronomy revival, Middle East space programs}
}

\begin{document}
\maketitle

\begin{abstract}
From al-Sufi's tenth-century observation of the Andromeda Galaxy as a "little cloud" to contemporary space missions, Islamic astronomy represents a millennium-spanning tradition of innovation and knowledge. This study traces its trajectory through three phases: the Golden Age (8th to 15th centuries), when scholars such as al-Biruni, al-Battani, and Ibn Sina developed instruments, cataloged the heavens, and refined theories that later influenced Copernicus; a period of decline (late 15th to 17th centuries), shaped by political fragmentation, economic shifts, and the delayed adoption of technologies such as printing and the telescope; and today's revival, marked by observatory collaborations, Olympiad successes, and emerging space programs in Morocco, Iran, Turkey, the UAE, and Saudi Arabia. This comparative analysis with Chinese and European scientific traditions shows how Islamic astronomy provided a vital link in the global history of science, transmitting mathematical rigor, observational methods, and Arabic star names that are still used today. The contemporary resurgence signals the potential for renewed contributions to astrophysics, provided that it is supported by regional observatory networks, space-based research initiatives, and educational frameworks that integrate historical heritage with modern computational science.
\end{abstract}


\section{Introduction: Looking Up, Looking Forward}
This paper traces the trajectory of astronomy in the Islamic world: the pioneering work of early Muslim astronomers who mapped the heavens, the subsequent period of decline, and the revival that has begun in the modern era.  

In 964 CE, astronomer Abd al-Rahman al-Sufi looked into the night sky of the Persian city of Isfahan and recorded something extraordinary. Adjacent to the familiar stars of the Andromeda constellation, he noticed an obscure smudge of light, a "little cloud" that seemed to drift in the dark night \cite{kunitzsch1989fixedstars}. With no access to telescopes, which had yet to be invented, and no knowledge of galaxies, he discovered one trillion suns, an island universe 2.5 million light-years away. His observation would make history, becoming the first written record of a galaxy outside the Milky Way. A millennium later, we are now able to revisit these very same celestial wonders, contrasting al-Sufi’s naked-eye accounts with new high-resolution images captured through modern astrophotography. 

A Persian scholar, Al-Battani, documenting the sky from a medieval courtyard, perfectly captured the essence of Islamic astronomy at its peak. For nearly seven centuries, the Islamic world advanced humanity’s understanding of the universe with unmatched institutional support and sophistication in mathematics. This period was the Golden Age for Islamic astronomy: faith and philosophy converged with empiricism to depict the heavens with precision that would not be surpassed until the 16th century. Al-Battani's planetary observations achieved accuracies that Copernicus would cite centuries later as a foundation of heliocentrism. Ibn Sina (Avicenna) recorded the changing colors and "sparks" of the Supernova 1006 "new star" (SN 1006) with unusual detail that scholars still analyze in modern research literature. Al-Biruni used the timings of lunar eclipses to determine longitude differences between cities and separately applied spherical trigonometry in a mountain horizon experiment to estimate Earth's radius; this was centuries before trigonometry was systematized in Europe by Regiomontanus.       

Their innovations extended beyond pure observation to technological and methodological breakthroughs that defined pre-telescopic astronomy. The astrolabe, an ancient instrument used for navigation, became both a scientific instrument and a spiritual compass, guiding prayer times and navigation with equal precision across three continents. Massive mural instruments, culminating in sextants nearly 40 meters in radius at Samarkand, pushed arcminute-level scale divisions with multi-arcminute accuracy in practice. That standard was passed only in the late sixteenth century, at Tycho Brahe’s Uraniborg on Hven (constructed 1576 to 1580), where giant naked-eye instruments yielded catalog accuracies of about 1 to 2 arcminutes. Star catalogues systematically corrected ancient Greek errors while preserving observational data that remain scientifically valuable. This was a civilization where science and spirituality reinforced each other, and where measuring the heavens deepened faith rather than threaten it. 

However, the pace of progress slowed, and obstacles began to interfere; political fragmentation destroyed the support networks behind long-term exploratory projects, and Samarkand's great observatory was abandoned after Ulugh Beg's 1449 assassination. As Atlantic-oriented trade reshaped economies, Europe absorbed Islamic astronomy--Al-Battani's trigonometry was explicitly cited by Copernicus--then amplified it with the arrival of printing in the 15th century, the invention of the telescope in the Netherlands (1608),  its rapid astronomical use by Galileo (1609), and the formation of universities that helped sustain research in spite of numerous political upheavals.  

In recent years, however, astronomical progress in the region has started to gain momentum again. Across many nations in the Middle East, scientists, students, and educational institutions are re-engaging with astronomy in technically advanced ways. Moroccan observatories are contributing valuable data to global sky surveys. In Iran, students trained in computational astrophysics are winning International Astronomy Olympiads, while Saudi astronomers have developed ways to improve imaging lunar crescents even in the daytime. The UAE now operates the Hope probe orbiting Mars, a mission that has received the attention of the public. These achievements are representations of the resurgence of astronomical research in the Islamic world and are inspired, in a way, by a centuries-old legacy in the region. 

In the section that follows, we review categories of celestial objects studied by scholars in the Islamic world, such as al-Sufi's little cloudy mass in constellation Andromeda, comets, supernovae such as the 1006 and 1054 events, and solar and lunar eclipse records that bound theory to practice. Next, we compare these observations using Chinese and European traditions to show how scientific methods, institutions, and knowledge transitioned between cultures. Then we turn our attention to the instruments and calculation tools such as astrolabes, sine quadrants, mural quadrants, and giant sextants, as well as observatories from Maragha to Samarkand and show how these advances informed early modern Europe. Most importantly, we investigate the causes of later decline, including political fragmentation, economic realignment, shift in scientific and religious thinking, and slower institutional incorporation of printing and the telescope in the Islamic world. Finally, we consider signs of revitalization, such as observatory collaborations, student competitions in Olympiads, and national space ambitions in Morocco, Iran, Turkey, the UAE, and Saudi Arabia. We also make practical recommendations that follow this history. This includes regional observatory networks, domestic participation in international space and survey missions, and educational programs that involve historical heritage and modern computational science.

\section{Celestial Objects in the Golden Age}
\subsection{Galaxies and Other Deep Sky Objects}
\begin{figure}[h]
    \centering
    \includegraphics[width=0.9\linewidth]{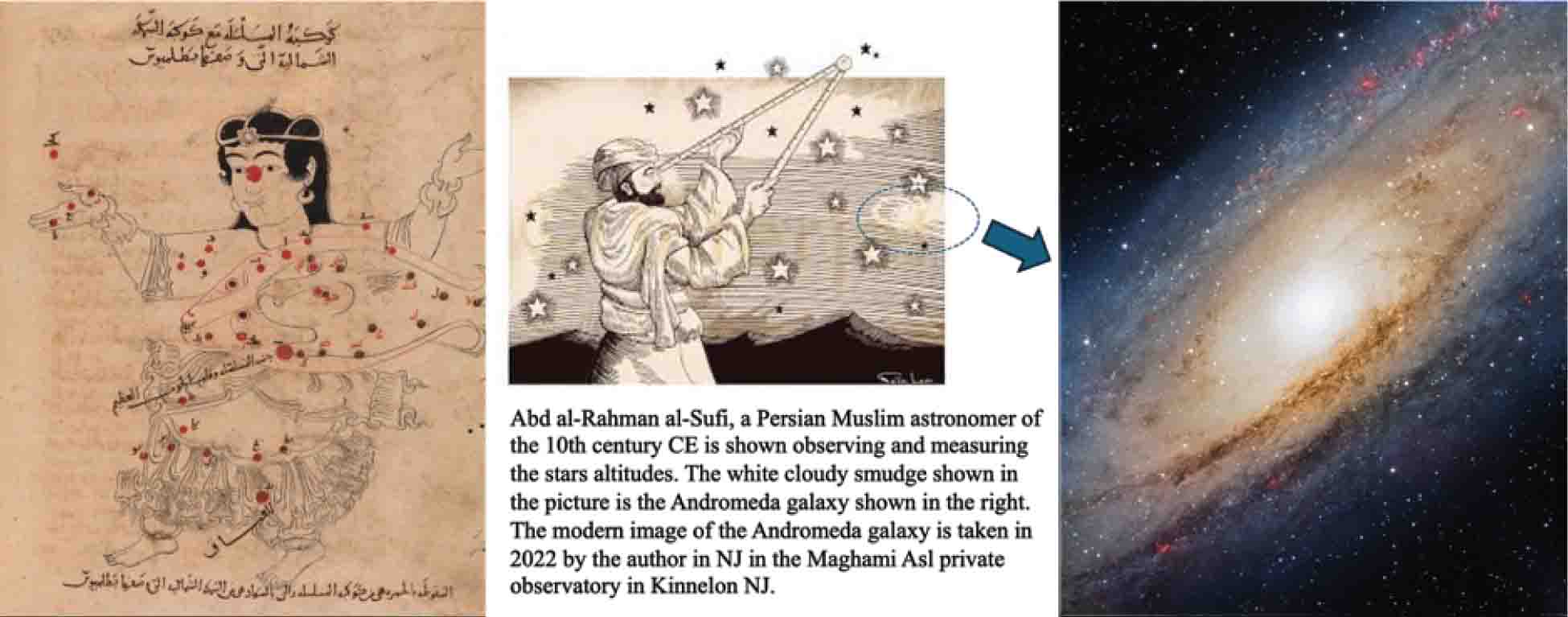}
    \caption{Left and middle: al-Sufi's manuscript depiction of Andromeda in the \textit{Book of Fixed Stars} (964 CE), where he described a faint smudge of light as “al-latka al-bayda,” the “little cloud.” Right: a modern CCD image of the Andromeda Galaxy, taken in 2022 from Kinnelon, New Jersey.}
    \label{fig:andromeda}
\end{figure}

Centuries before the telescopic age, the Muslim astronomer \textbf{Abd al-Rahman al-Sufi} (903–986) charted the heavens with such precision that he left humanity's first written record of what we now know to be a galaxy beyond the Milky Way. Living in Isfahan during the Buwayhid dynasty and the Persian Renaissance, al-Sufi completed his celebrated \textit{Book of Fixed Stars} in 964 CE. The work not only corrected Ptolemy's Greek star charts but also introduced new observations based on his own naked-eye studies. Among the most remarkable was his description of a faint patch of light in the constellation Andromeda, which he called \textit{al-latkha al-sahabiya} (“the little cloud”). He had no way of knowing that this smudge contained a trillion stars, an island universe 2.5 million light-years away \citep{gingerich1986islamicrole, north2008cosmos}. Yet his careful description and illustration (Figure~\ref{fig:andromeda}) preserve the earliest account of the Andromeda Galaxy (M31).

Al-Sufi’s description of the Andromeda Galaxy reflects a crucial cultural dimension often overlooked in astronomical histories. Where Greek astronomers saw the chained princess Andromeda, Islamic-era observers identified two fish constellations (Figure~\ref{fig:andromeda}). Al-Sufi described his “nebulous smear” as positioned “before the mouth of a Big Fish” (\textit{al-hut al-kabir}), demonstrating how different cultural frameworks shaped the same celestial observations. Comparing his sketch with a modern CCD image taken in 2022 from a private observatory in Kinnelon, New Jersey, demonstrates not only the continuity of human curiosity but also the extraordinary acuity of pre-telescopic observers.

\begin{figure}[h]
    \centering
    \includegraphics[width=0.5\linewidth]{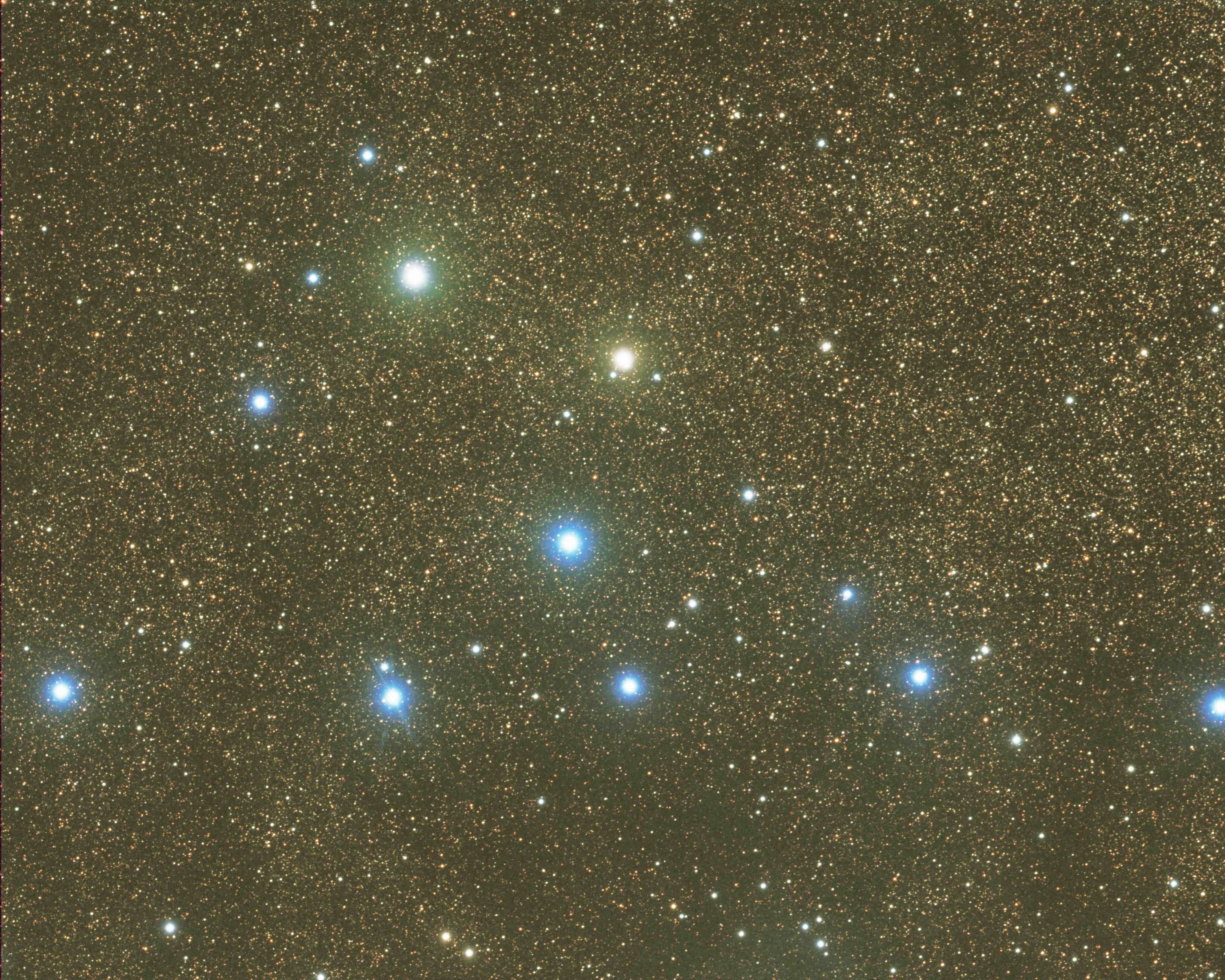}
    \caption{Al-Sufi Cluster, also known as Brocchi's Cluster, and Collinder 399 (CR 399). The star system is located at constellation Vulpecula. It is also known by the nickname \textit{Coathanger}. This modern day astrophotograph was taken by the author in 2025 from the family private Deep Space Observatory in Fort Davis, Texas.}
    \label{fig:cr399}
\end{figure}

Al-Sufi’s observational program extended far beyond Andromeda. He systematically catalogued other nebulous objects, including the Omicron Velorum star cluster as a “nebulous star” and an additional “nebulous object” in Vulpecula, now known as al-Sufi’s Cluster (CR 399 or Brocchi’s Cluster) \citep{gingerich1986islamicrole}. A modern era CCD image of al-Sufi's cluster in RGB channels taken by the author is shown in Figure~\ref{fig:cr399}. While he did not conceive of these as the distant stellar formations we understand today, his catalog of nebulous objects preserved critical evidence of structures later central to modern astrophysics. He recorded travelers’ reports of the Magellanic Clouds—irregular galaxies visible only from the Southern Hemisphere \citep{gingerich1986islamicrole}. 

\begin{figure}[h]
    \centering
    \includegraphics[width=0.8\linewidth]{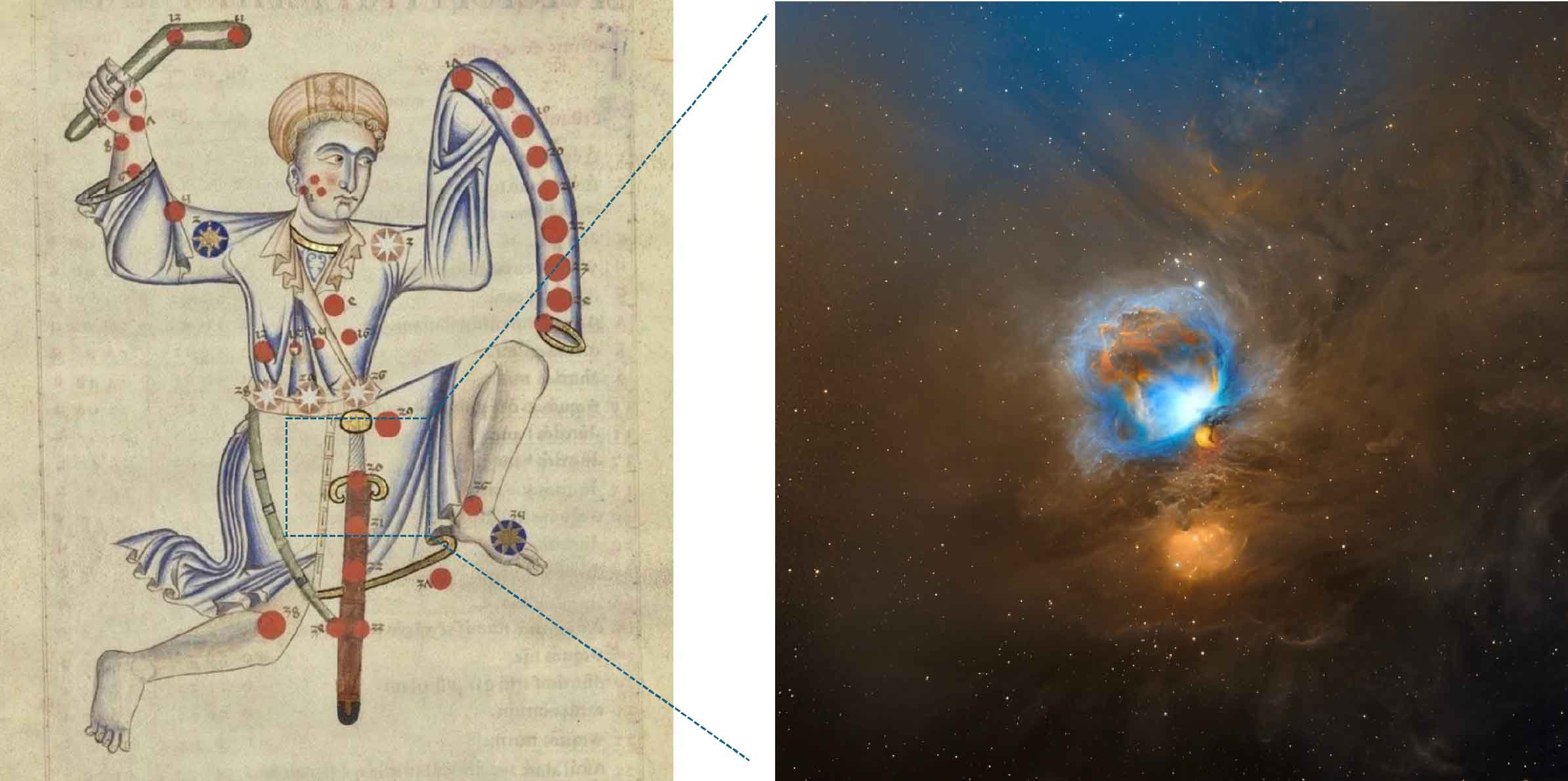}
    \caption{Left: the Orion constellation in a Latin translation of al-Sufi's \textit{Book of Fixed Stars} (Bologna, 1250–1275). Right: a modern astrophotograph of the Orion Nebula taken by the author in 2021 from Kinnelon, New Jersey.}
    \label{fig:orion}
\end{figure}

\begin{figure}[h]
    \centering
    \includegraphics[width=0.8\linewidth]{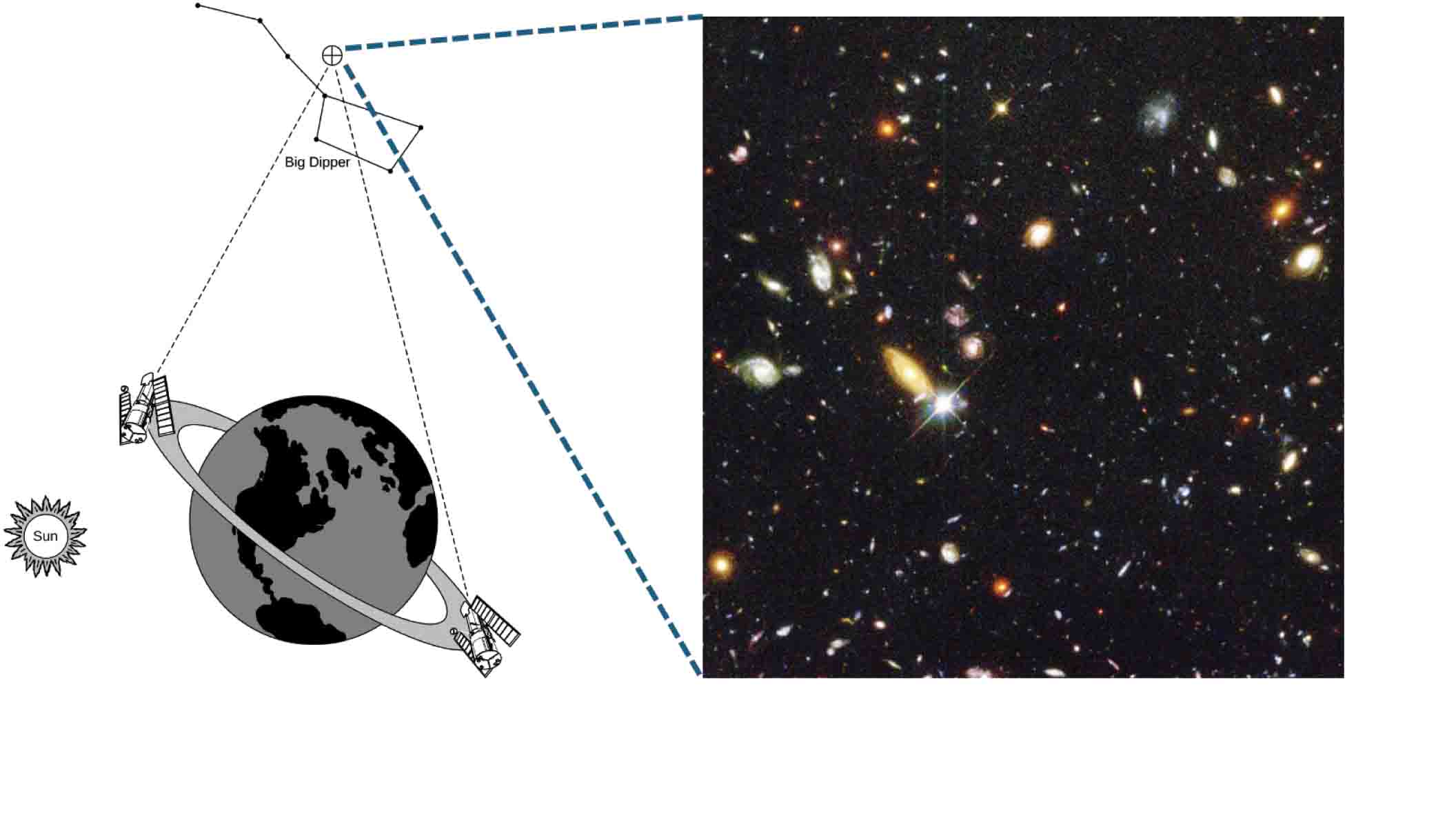}
    \caption{The Hubble Deep Field (1995), revealing nearly 3,000 galaxies in a patch of sky just 1/13th the diameter of the full Moon (NASA/ESA/STScI).}
    \label{fig:hubble_field}
\end{figure}

Al-Sufi’s work was also translated into Latin at the University of Bologna in 1250. One of the most notable illustrations from this manuscript depicts the Orion constellation (Figure~\ref{fig:orion}). It remains unclear whether al-Sufi noted the haziness in Orion's sword, a prominent nebulous object later identified as the Orion Nebula. This empirical rigor distinguished astronomy in the Islamic world from many contemporary approaches. Where medieval European astronomy often emphasized theoretical synthesis of inherited texts, scholars like al-Sufi prioritized direct observation and empirical correction of classical sources \citep{burnett2009toledo}. His work exemplifies a scientific culture that verified inherited knowledge through observation, transforming wonder into measurement while integrating astronomy into religious, navigational, and cultural life.
The comparative significance of this achievement becomes clear when examining Islamic and European treatments of the Andromeda Galaxy before the telescope's invention. As shown in Table~\ref{tab:andromeda_comparison}, al-Sufi's description from 964 CE was centuries ahead of Europe, where Andromeda was not independently recognized until the telescopic age. In 1612, the German astronomer Simon Marius provided the first telescopic description of Andromeda, describing its central glow fading outward. In 1764, Charles Messier catalogued it as M31, still assuming it was a nebula within the Milky Way \citep{north2008cosmos}. The true breakthrough came in 1924, when Edwin Hubble used Cepheid variable stars to prove that Andromeda lay millions of light-years away, revealing galaxies as separate “island universes” and transforming cosmology \citep{hubble1925cepheids}. A striking illustration of this new understanding is Figure~\ref{fig:hubble_field}, the Hubble Deep Field, which revealed thousands of galaxies in various stages of their evolution.
The cultural memory of al-Sufi's achievement endured. His constellation drawings circulated in European manuscripts after Arabic-to-Latin translations in Toledo and Sicily, shaping both scientific and artistic traditions \citep{burnett2009toledo}. In modern times, the Astronomy Society of Iran has honored his legacy through international “Sufi Observing Competitions,” linking today’s amateur astronomers to the tradition he pioneered \citep{brookings2024meSpace}.

Today, Andromeda remains a central focus of astrophysics as both our nearest large neighbor and a future merger partner of the Milky Way in about 4.5 billion years \citep{vandermarel2012m31vector}. From al-Sufi’s courtyard in Isfahan to space-based observatories, the thread of observation is continuous. His “little cloud” represents more than a naked-eye discovery; it exemplifies how empirical observation, cultural integration, and cross-civilizational transmission laid the foundations of our understanding of galaxies.

\begin{table}[h]
\centering
\caption{Comparing Islamic and European Descriptions of the Andromeda Galaxy Before the Telescope.}
\begin{tabular}{p{4cm} p{5cm} p{5cm}}
\toprule
\textbf{Aspect} & \textbf{Islamic World (al-Sufi, 964 CE)} & \textbf{Medieval Europe (Pre-1600)} \\
\midrule
Observational Approach & Direct naked-eye observation; systematic star chart revision & Reliance on Greek/Arabic texts; limited new observation \\
Description & “Little cloud” (\textit{al-latka al-bayda}) with accurate placement & No independent record; dependent on translations \\
Scientific Method & Empirical correction of inherited sources & Preservation and commentary on texts \\
Cultural Integration & Connected to prayer times, calendars, navigation & Primarily academic, detached from daily use \\
Legacy Impact & Direct transmission via translations, influencing Renaissance astronomy & Received Islamic knowledge, prepared for telescopic age \\
\bottomrule
\end{tabular}
\label{tab:andromeda_comparison}
\end{table}

\subsection{Supernovae in Chinese and Islamic Records}

In 1006 CE, a brilliant new light appeared in the sky—visible across the Islamic world, China, Japan, and Europe—so luminous that it could even be seen during the day. This celestial event, later identified as \textbf{Supernova SN 1006}, was the brightest recorded supernova in human history. The Egyptian astronomer \textbf{Ali ibn Ridwan} provided one of the most detailed eyewitness accounts, describing it as a luminous body “two and a half to three times the size of Venus, with light bright as that of a quarter of the Moon.” He noted that the object persisted for months, gradually fading after more than a year. Modern astrophysicists have used his careful descriptions—particularly his observations of brightness and duration—to determine that SN 1006 was a Type Ia supernova located roughly 7,200 light-years away in the constellation Lupus \citep{green2003historicalsn}.
\begin{figure}[h]
    \centering
    \includegraphics[width=0.9\linewidth]{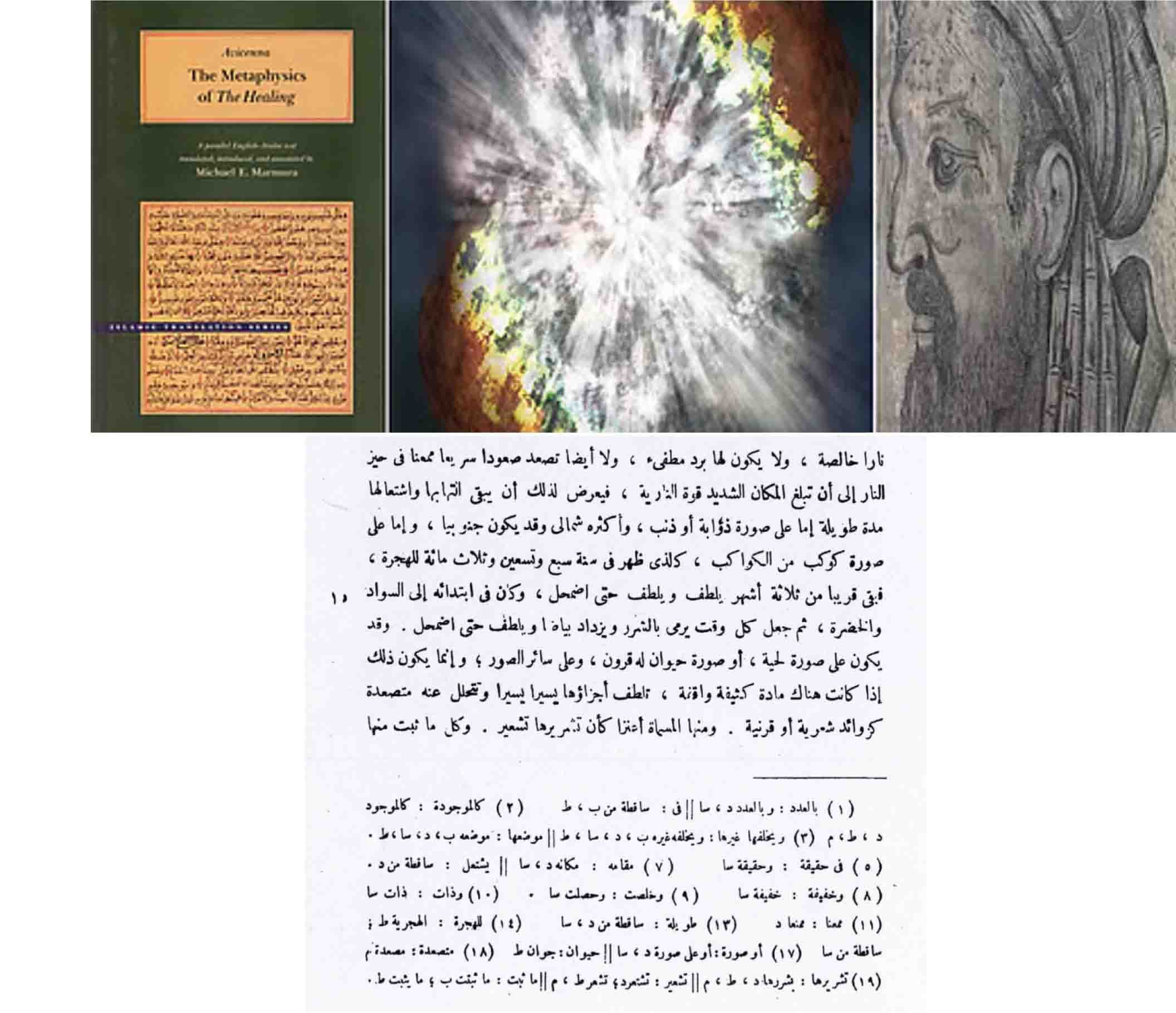}
    \caption{Ibn Sina's description of SN 1006 from \textit{Kitab al-Shifa} (\textit{The Book of Healing}), showing his detailed account of the supernova’s changing appearance over time. This work exemplifies the integration of astronomy within philosophical and scientific traditions of the Islamic Golden Age.}
    \label{fig:ibn_sina}
\end{figure}
The polymath \textbf{Ibn Sina (Avicenna)} (980–1037), one of the foremost thinkers of the Islamic Golden Age, also recorded this event in his scientific-philosophical encyclopedia \textit{Kitab al-Shifa} (The Book of Healing). A modern study by Neuhauser, Ehrig-Eggert, and Kunitzsch translated Ibn Sina’s detailed description of this celestial phenomenon \citep{neuhauser2013sn1006}. He wrote:

\begin{quote}
“It therefore happens that the burning and flaming stays for a (long) while, either in the form of a lock of hair or with a tail, mostly in the north but sometimes also in the south, or in the form of a star among the stars [\textit{kawkab min al-kawakib}] like the one which appeared in the year 397. It remained for close to three months, getting fainter and fainter until it disappeared; at the beginning it was towards a darkness and greenness, then it began to throw out sparks [\textit{yarmi bi-l-sharar}] all the time, and then it became more and more white, then became fainter and disappeared.”
\end{quote}

This record is exceptional for providing qualitative details—color, texture, and duration—beyond simple brightness. Ibn Sina noted that the object transitioned from “darkness and greenness” to brilliance before fading, a description consistent with the color evolution of a supernova. These observations enhance modern reconstructions of SN 1006, which reached an apparent magnitude of negative 7.5, bright enough to cast shadows at night \citep{schaefer1996visibilitysn}.

The event’s timing, between April 30 and May 1 of 1006 CE, and its location in Lupus, align precisely with modern astrophysical data \citep{green2003historicalsn}. Parallel Chinese records documented similar details, confirming its brightness and long visibility. Together, these independent records illustrate how Islamic and East Asian scholars produced complementary datasets, strengthening the global history of pre-telescopic astronomy.

A generation later, in 1054 CE, another “new star” appeared in Taurus, now identified as \textbf{Supernova SN 1054}—the progenitor of the Crab Nebula (M1). While the most precise positional data came from Chinese astronomers, Islamic observers also recorded the event. Chroniclers in Baghdad and Constantinople, including the physician \textbf{Ibn Butlan}, reported the sudden appearance of the star and noted its persistence for many months. These multiple accounts demonstrate the high level of astronomical literacy and observation within both Islamic and Chinese traditions \citep{chapman1984sn1054}.

\begin{figure}[h]
    \centering
    \includegraphics[width=0.9\linewidth]{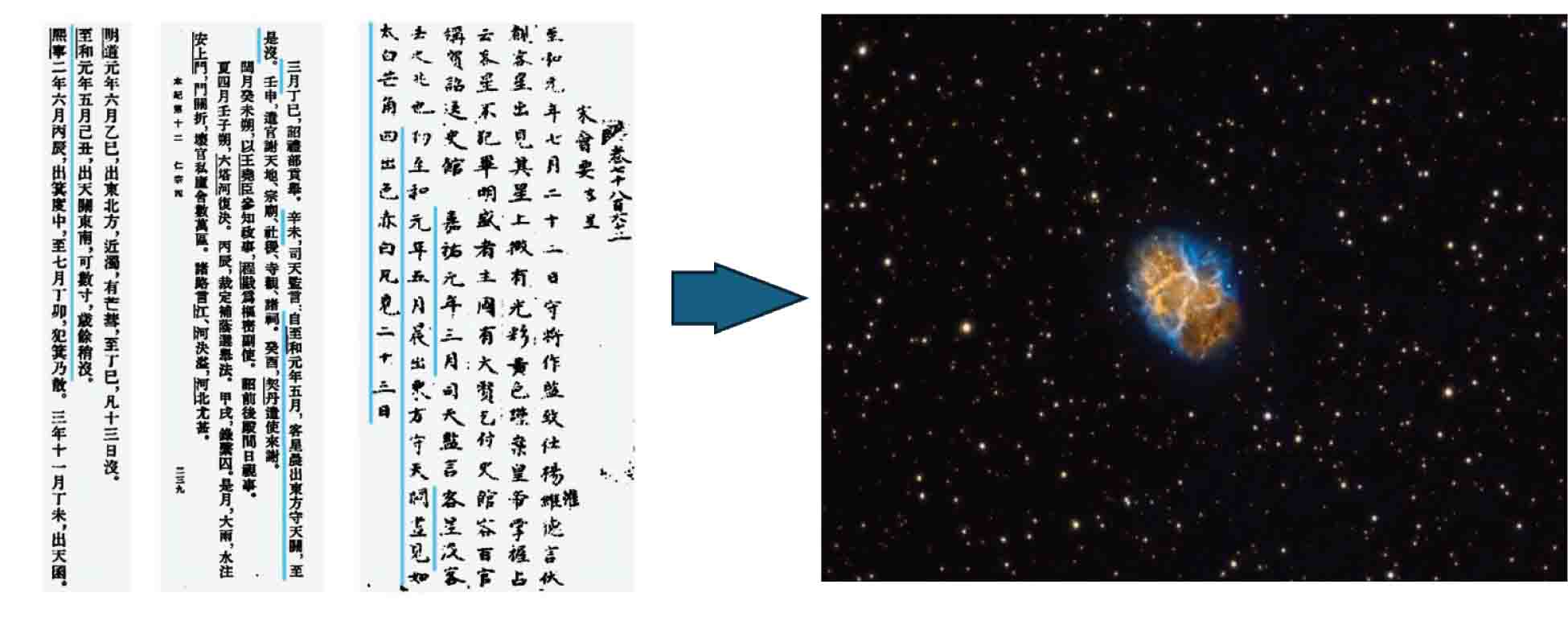}
    \caption{Left: Chinese records of SN 1054 from historical documents. Right: Modern image of the Crab Nebula (supernova remnant SN 1054), taken in 2021 from the Maghami Asl Private Observatory in Kinnelon, New Jersey. Together, these sources illustrate the global documentation of one cosmic event.}
    \label{fig:crab}
\end{figure}

Islamic chroniclers’ qualitative reports of SN 1054 complement the quantitative precision of Chinese measurements, which were accurate enough to locate the Crab Nebula’s remnants today (Figure~\ref{fig:crab}). This cross-civilizational collaboration underscores the global continuity of astronomical practice.

Since \textbf{Kepler’s Supernova} of 1604—the most recent visible in our galaxy—all subsequent supernovae observed have been extragalactic. The continuity of observation from medieval scholars to modern astrophysics illustrates the enduring scientific value of these transient cosmic phenomena. 

\begin{figure}[h]
    \centering
    \includegraphics[width=0.67\linewidth]{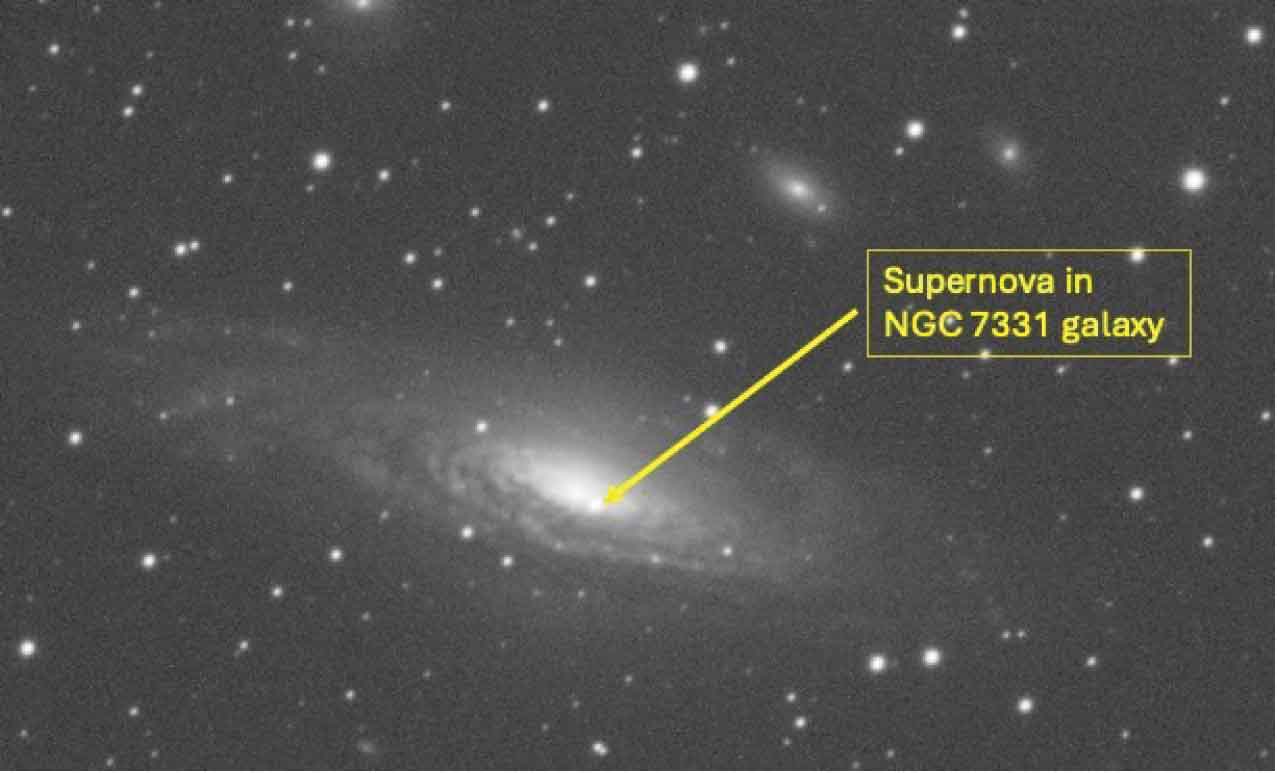}
    \caption{Supernova SN 2025rbs in galaxy NGC 7331, captured in 2025 from the Maghami Asl Private Observatory in Kinnelon, New Jersey. This Type Ia supernova connects contemporary observation to the same cosmic processes chronicled by Ibn Sina and Ibn Ridwan.}
    \label{fig:sn2025rbs}
\end{figure}

A recent example, \textbf{Supernova SN 2025rbs}, discovered in 2025 in the spiral galaxy NGC 7331 demonstrates this continuity. Figure~\ref{fig:sn2025rbs} shows the image of this supernova taken by the author just a few days after its discovery in July of 2025. Galaxy NGC 7331, shown in the figured was cataloged by William Herschel in 1784. It lies 44 million light-years away in Pegasus and is structurally similar to both Andromeda and the Milky Way. SN 2025rbs is the first Type Ia supernova observed in this galaxy; all previous NGC 7331 supernovae were Type II. Type Ia supernovae serve as \textit{standard candles}—cosmic distance markers of uniform luminosity. Their use enabled the discovery of the universe’s accelerating expansion, earning the 2011 Nobel Prize in Physics for Perlmutter, Schmidt, and Riess \citep{riess1998acceluniverse}.
The tradition of meticulous supernova observation—exemplified by Ibn Sina’s description in \textit{The Book of Healing}—thus links medieval Islamic science to the foundations of modern cosmology. Unlike his purely medical \textit{Canon of Medicine}, this astronomical record appears in a philosophical encyclopedia, reflecting the Islamic Golden Age’s integrative approach to knowledge \citep{gutas2001avicenna}. Astronomy, medicine, philosophy, and theology coexisted as parts of a unified intellectual pursuit. 

For scholars such as Ibn Sina, al-Biruni, and al-Farabi, celestial observation was both a practical science (for prayer times, navigation, and calendars) and a philosophical inquiry into cosmic order. Ibn Sina’s record of SN 1006 exemplifies this synthesis—where empirical precision and metaphysical reflection met in pursuit of understanding the universe.

From Ibn Ridwan’s 1006 chronicle to modern images of SN 2025rbs, the observation of stellar death has remained a bridge across centuries. These events illuminate not only the life cycles of stars but also humanity’s continuous effort to interpret the heavens through science, philosophy, and shared wonder.

\subsection{Comets During the Islamic Golden Age}
Comets are among the most striking and sometimes ominous celestial phenomena humans can witness. Medieval Islamic historians and astronomers referred to them as “tailed stars,” appearing suddenly and streaking across the heavens with fiery trains that seemed to herald cosmic change. These visitors inspired both scientific scrutiny and poetic reflection. Chroniclers recorded not only their appearances but also details such as tail length, color, direction of motion, and duration of visibility.

In Arabic, comets were often termed \textit{kawkab dhu-dhanab} (“a star with a tail”) or \textit{kawkab dhu-dhanab sha’r} (“a star with a lock of hair”), while the term \textit{nayzak} (“spear”) captured both their brilliance and suddenness—though it was sometimes also applied to meteors \citep{cook1999cometsmeteors, rada1999catalogue}. What distinguished comets most clearly from meteors was their persistence. Unlike the fleeting streaks of shooting stars, comets lingered for weeks or months, sometimes rivaling planetary brightness and leaving luminous arcs across the night sky.

The most famous apparition of the Islamic Golden Age was \textbf{Halley’s Comet in 837–838 CE}, when it passed closer to Earth than at any other recorded time. In Baghdad—the intellectual capital of the Abbasid world—observers marveled at a bright star with a sweeping tail stretching across the sky. Fragments of a lost treatise by the philosopher \textbf{al-Kindī} (801–873) indicate he recorded this comet systematically, treating it as a scientific object rather than an omen \citep{yazdi2014alkindi837}. Thinkers such as \textbf{Ibn Sina} and \textbf{al-Biruni} debated its nature, rejecting Aristotle’s vapor theory and arguing instead for celestial origins—an intellectual shift that deepened empirical approaches to cosmic phenomena.
\begin{figure}[h]
    \centering
    \includegraphics[width=0.75\linewidth]{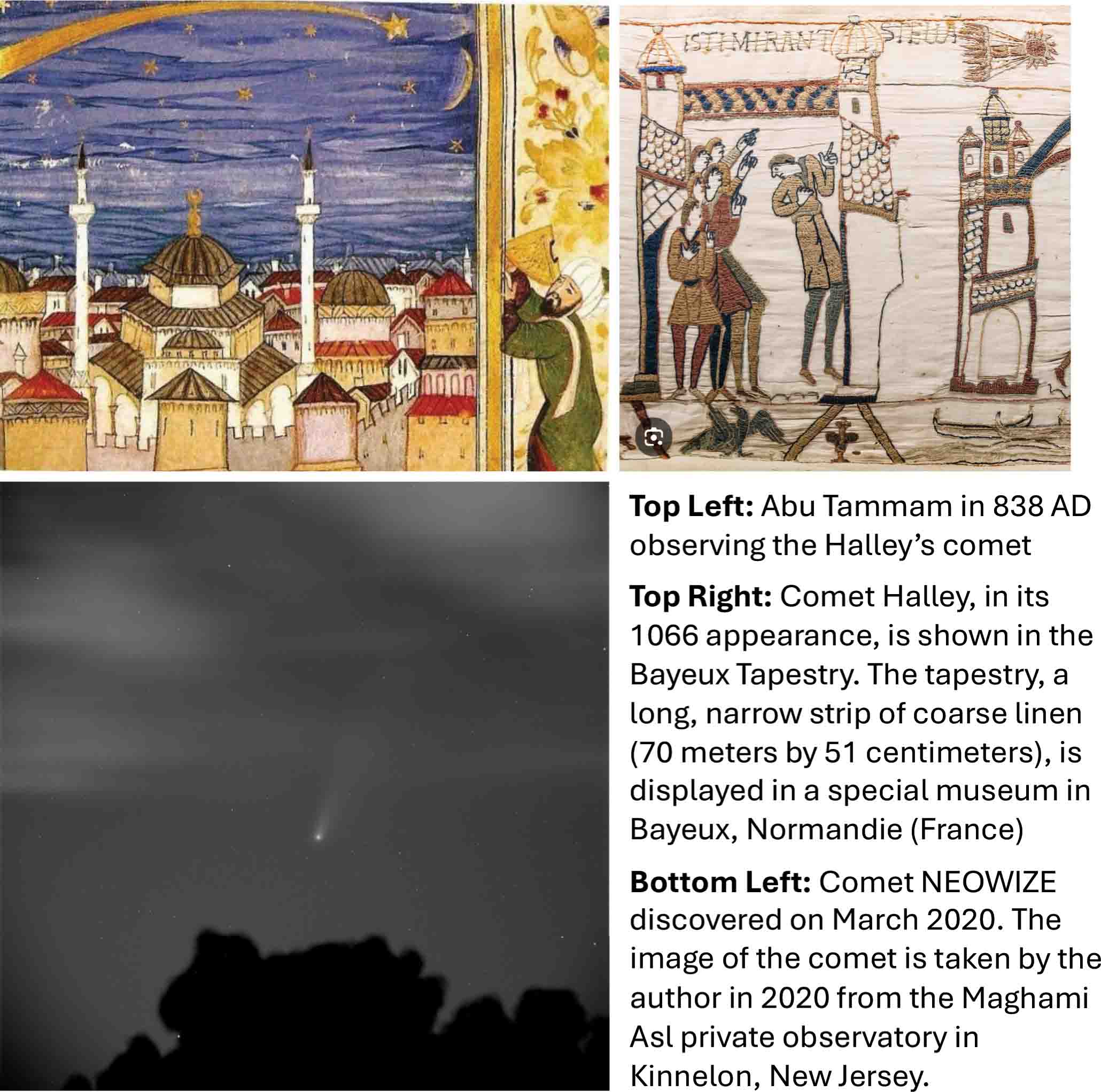}
    \caption{Left: Abu Tammam observing Halley’s Comet in 838 CE during the Islamic Golden Age, representing the union of scientific observation and poetic reflection. Center: Halley’s Comet in 1066 CE depicted in the Bayeux Tapestry as an omen of conquest. Right: Comet NEOWISE photographed in 2020 from the Maghami Asl Private Observatory in Kinnelon, New Jersey, continuing the tradition of systematic cometary observation.}
    \label{fig:comets}
\end{figure}
The same comet also left its mark on literature. The renowned poet \textbf{Abu Tammam} (804–845 CE) described a “western star with a tail,” widely regarded as a poetic record of Halley’s Comet \citep{cook2002cometspoetry}. His verse exemplifies how comets inspired wonder, symbolism, and reflection on fate, turning astronomical observation into cultural expression. This interplay between poetry and astronomy demonstrates how deeply comets entered both scientific and artistic consciousness during the Islamic Golden Age.

From al-Andalus in the west to Khurasan in the east, Muslim chroniclers recorded cometary appearances with exceptional precision. A typical entry might read: “In the month of Ramadan, a bright star with a long tail rose in the east, moving nightly toward the south until it disappeared.” Such attention to timing, direction, and brightness enables modern scholars to identify historical cometary events, including multiple returns of Halley’s Comet \citep{cook1999cometsmeteors, kronk1999cometography}. 

The return of Halley’s Comet in \textbf{1066 CE}—immortalized in Europe on the Bayeux Tapestry as an omen of the Norman Conquest—was also recorded by Muslim historians in Iraq and Syria (Figure~\ref{fig:comets}). They described it as a bright tailed star visible for several weeks, integrating it into a cosmological worldview where celestial and political events reflected divine order \citep{stephenson2003earthrotation}. A later reappearance in \textbf{1222 CE} inspired vivid descriptions in Damascus: “A great comet appeared in the east, with a large and thick tail visible for ten nights, shifting nightly until it vanished at the end of Ramadan.” For societies already shaken by Mongol invasions, the comet carried dual meanings—a marvel of divine creation and an omen of upheaval \citep{peacock2019mongolanatolia, cook1999cometsmeteors}.

Despite their symbolic significance, Muslim astronomers combined cultural interpretation with rigorous observation. Some accounts note color transitions—reddish at first, then whitening—while others trace motion against fixed stars. Cross-referencing these with Chinese and European sources allows modern astronomers to reconstruct the orbits of historical comets, particularly Halley’s periodic returns \citep{stephenson1988eastasian, kronk1999cometography}. Yet unlike European chroniclers, who often saw comets as divine warnings, Islamic observers viewed them through a framework balancing theology and empirical inquiry.

Philosophers and scientists continued to debate cometary origins. \textbf{Ibn Sina} and \textbf{al-Biruni}, drawing on and modifying Aristotelian ideas, questioned whether comets were atmospheric vapors or true celestial bodies. By the 13th century, at the \textbf{Maragha Observatory}, \textbf{Nasir al-Din al-Tusi} (1201–1274) argued that astronomy must be liberated from rigid metaphysics and grounded in mathematics and observation. A century later, \textbf{Ali Qushji} (d. 1474) advanced this view, asserting that astronomy should stand as an independent empirical science \citep{ragep2001freeing}. Within this evolving intellectual climate, comets were redefined—not as omens or “exhalations,” but as genuine celestial entities demanding quantitative explanation.

The Islamic-era record of cometary observation reveals a unique synthesis of empirical precision and cultural meaning. Scholars documented brightness, color, tail length, and direction with striking accuracy, providing modern astrophysicists valuable data for orbit reconstruction \citep{cook1999cometsmeteors, rada1999catalogue}. 

This tradition extends to the present. In 2020, \textbf{Comet NEOWISE (C/2020 F3)} illuminated northern skies, the brightest since Hale-Bopp in 1997 and Halley’s 1986 return. Figure~\ref{fig:comets} shows a photo of comet NEOWIZE taken by the author in 2020. While Halley’s Comet revisits every 76 years, NEOWISE follows an immense 6,800-year orbit—its last perihelion occurring in 4772 BCE and its next not due until 8812 CE \citep{nasa2020neowise}. 

From Abu Tammam’s poetic vision of Halley’s Comet in 838 CE to modern astrophotographs of NEOWISE, comets remain both scientific objects and symbols of wonder. For scholars of the Islamic Golden Age, they offered opportunities to unite observation and meaning; for modern scientists, they remain messengers from the solar system’s infancy, preserving material from its earliest epochs.

While Islamic astronomers integrated systematic observation with philosophical interpretation, most medieval European chronicles portrayed comets as divine warnings—signs of plague, famine, or war. The Anglo-Saxon Chronicle described Halley’s Comet in 1066 as a portent of disaster, while Chinese astronomers recorded the same event with bureaucratic precision as part of state celestial monitoring. Only in the Renaissance, through \textbf{Tycho Brahe}’s meticulous measurements of the 1577 comet, was it definitively shown that comets were celestial rather than atmospheric phenomena \citep{dreyer2014tycho}. 

The comparative study of these traditions highlights how differing intellectual and cultural frameworks shaped humanity’s understanding of comets. Islamic astronomers occupied a pivotal position between symbolic interpretation and mathematical empiricism—laying groundwork for the observational astronomy that continues today.

\subsection{Eclipses: Observation, Prediction, and Practice}

When the Sun dimmed at midday or the full Moon slipped into a red shadow at night, observers across the medieval world responded with awe and inquiry. In Arabic, solar and lunar eclipses were distinguished as \textit{al-kusūf} (solar) and \textit{al-khusūf} (lunar). From the earliest Islamic period, eclipses were framed as occasions for reflection rather than omens; the Prophet Muhammad clarified that the Sun and Moon do not eclipse for births or deaths and urged observance of \textit{salāt al-kusūf}, establishing a rational and devotional stance toward these events \citep[for historical discussions of medieval Islamic eclipse practice and records, see][]{stephenson1996eclipses}.
\begin{figure}[h]
    \centering
    \includegraphics[width=0.95\linewidth]{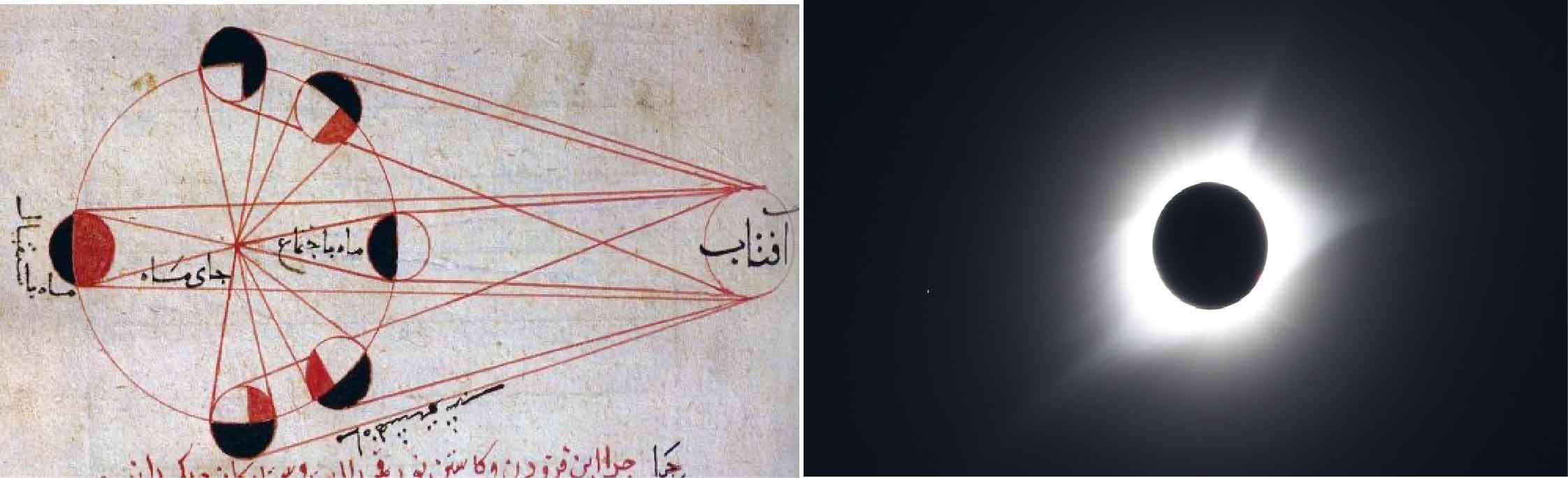}
    \caption{Left: Phases of the Moon in a Persian translation of al-Bīrūnī’s \textit{Kitāb al-Tafhīm} (Library of the Parliament, Tehran). Right: The solar eclipse of 21 August 2017 photographed in Jackson, Wyoming. The pairing illustrates continuity from medieval Islamic eclipse observation to modern astrophotography.}
    \label{fig:biruni_eclipse}
\end{figure}
\begin{figure}[h]
    \centering
    \includegraphics[width=0.6\linewidth]{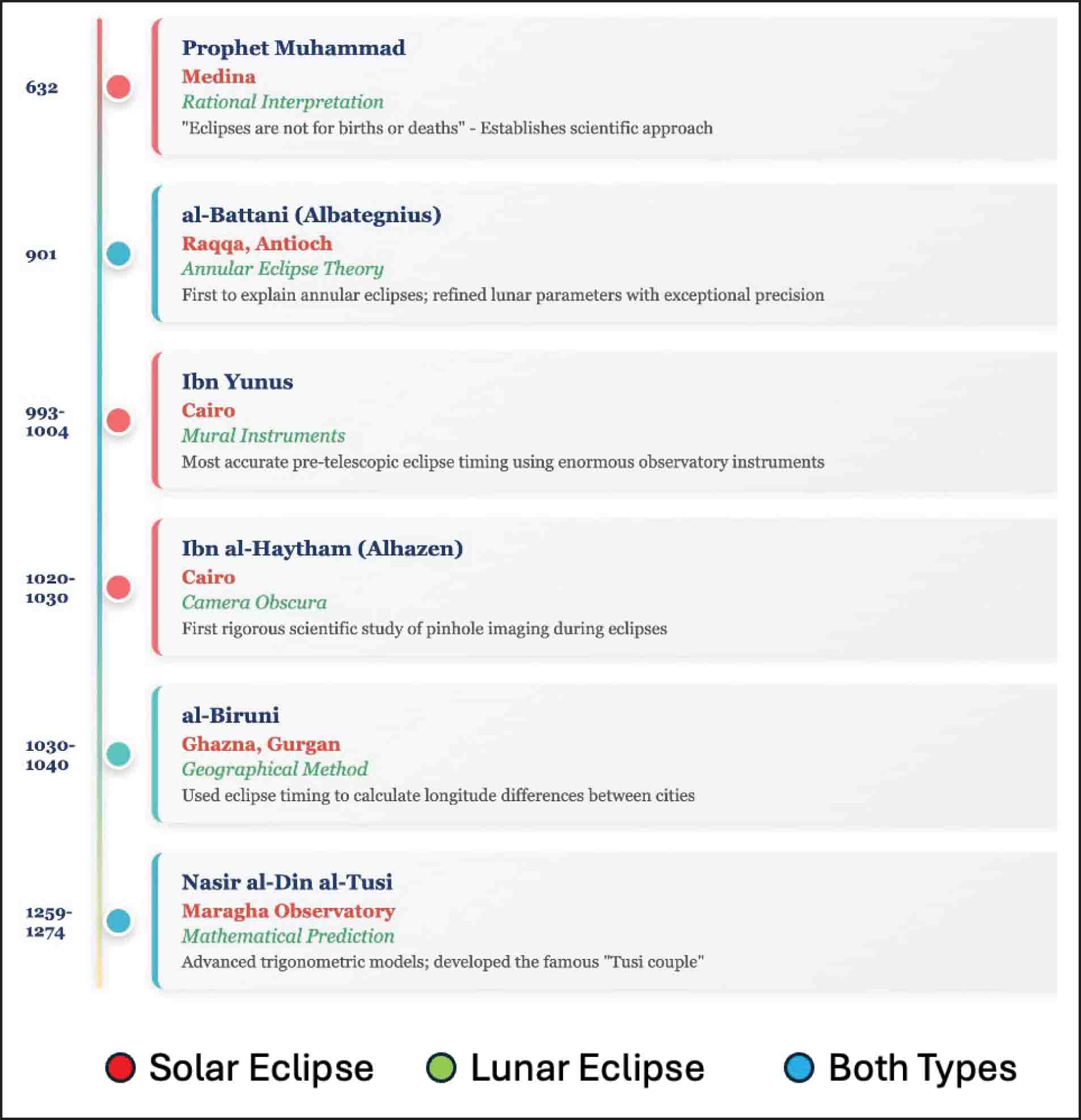}
    \caption{Timeline of eclipse observations in Islamic astronomy (632–1300 CE), showing the progression from early normative guidance to predictive \textit{zij} tables and observatory programs. Color coding distinguishes solar eclipses (red), lunar eclipses (teal), and scholars active on both (blue). Geographic spread from Medina to Marāgha indicates institutional continuity and cross-regional collaboration.}
    \label{fig:eclipse_timeline}
\end{figure}
By the ninth century, Islamic astronomers incorporated eclipses into \textit{zij} astronomical handbooks, enabling accurate predictions. Working in Raqqa and Antioch, \textbf{al-Battani} (858–929) observed the paired eclipses of 901 with exceptional precision, refined lunar parameters, and clearly explained annular eclipses via the varying apparent solar diameter with distance. His eclipse observations influenced both Islamic and European astronomy and remain valuable for modern studies of lunar motion \citep{stephenson1991precision}. Al-Battani's \textit{Zij al-Sabi} provided improved values for the solar year, obliquity of the ecliptic, and the Sun’s apogee, and clarified eclipse conditions; translated into Latin in the 12th century, it shaped European astronomy and was cited by Copernicus.

In Fatimid Cairo, \textbf{Ibn Yunus} (950–1009) compiled the monumental \textit{Zij al-Hakimi al-Kabir} using large mural instruments at al-Azhar. He recorded dozens of eclipses with careful timing of first contact and maximum phase, along with reference star altitudes. Centuries later, Western astronomers used his measurements to refine models of lunar motion \citep{stephenson1996eclipses}.

\textbf{Al-Biruni} (973–1050) transformed eclipses into tools of geography. In the \textit{Qanun al-Masudi} he explained how timing a lunar eclipse at different cities determines longitude differences, comparing observations at Gurgan and Ghazna with notable accuracy. His systematic documentation of lunar phases and eclipse geometry exemplifies the practical applications of Islamic astronomy and its continuity with modern work \citep{albiruni1954qanun}. Figure~\ref{fig:biruni_eclipse} juxtaposes a medieval phase illustration with a modern eclipse image.

By the 13th century, observatories such as Maragha consolidated eclipse prediction within broader reform programs. \textbf{Nasir-al-Din al-Tusi} (1201–1274) systematized planetary theory and improved trigonometric schemes for prediction. His “Tusi couple, ”generating linear motion from circular components, removed Ptolemy’s equant while preserving uniform motion, and later appears in Copernican models. Scholars of the Maragha tradition advanced the view that astronomy should stand independently of Aristotelian natural philosophy, grounded in mathematics and observation—an approach later articulated by Ali Qushji \citep{ragep2001freeing}.

Islamic era eclipse studies combined precise timing, positional astronomy, and methodological reflection. Figure~\ref{fig:eclipse_timeline} sketches the chronological development and geographical spread of systematic eclipse work from 632 to 1300, highlighting institutional continuity and evolving methods. Data from al Battani, Ibn Yunus, and al-Biruni remain historically significant for contemporary analyses of Earth’s rotation and lunar acceleration \citep{stephenson2003earthrotation}.

\section{Instruments of the Sky: Technology and Transmission}
Astronomy in the Islamic Golden Age joined theory to technology. Scholars designed instruments that turned observations of comets, eclipses, and planetary motions into measurements and predictions.

The most notable instrument was the astrolabe, a projection of the celestial sphere onto a flat plate. Building on Greek prototypes, Muslim scholars such as al-Fazari (8th c.) and al-sufi (10th c.) expanded designs and uses, including qibla finding and precise prayer times \citep{king1999instruments}. With an astrolabe, observers could find local time, measure stellar altitude, detect dawn, and determine the direction of Mecca. Thousands survive in museums, many with geometric decoration, indicating extensive use and cultural value \citep{morrison2007astrolabe}.

Alongside astrolabes, astronomers developed compact trigonometric tools. The sine quadrant (\textit{rub al-mujayyab}) enabled quick solutions for altitudes, prayer times, and qibla direction; other common devices included quadrants and armillary spheres used in observatories and teaching. These instruments later informed European navigation.

For high-precision work, astronomers built large fixed instruments. In 994 CE, Abū Maḥmūd al-Khujandī erected the Fakhri Sextant in Rayy, a 60° meridian arc for measuring the Sun’s solstitial altitude and deriving the obliquity and local latitude; later scholars, including al-Bīrūnī, refined the results \citep{albiruni1954qanun}. Such scale and accuracy foreshadowed European mural instruments, including Tycho Brahe’s quadrants \citep{dreyer2014tycho}.

By the 13th century, instrument-centered institutions emerged. The Maragha Observatory (founded 1259 under Hülegü; directed by Naṣir al-Din al-Tusi) housed mural quadrants, armillary spheres, celestial globes, instrument workshops, and an extensive library. There, Tusi formulated the “Tusi couple,” a mathematical device generating linear motion from uniform circular motions, removing Ptolemy’s equant while preserving uniformity; its kinematics appear in \textit{De Revolutionibus} \citep{ragep2001freeing}.

A century later, Ulugh Beg’s Samarkand Observatory extended this program. Its monumental sextant (radius \(\sim\)36 m) produced stellar positions more precise than Ptolemy’s and a star catalogue unsurpassed until Tycho \citep{north2008cosmos}. These instruments united mathematical theory with advanced engineering.
\begin{figure}[h]
  \centering
  \includegraphics[width=0.8\linewidth]{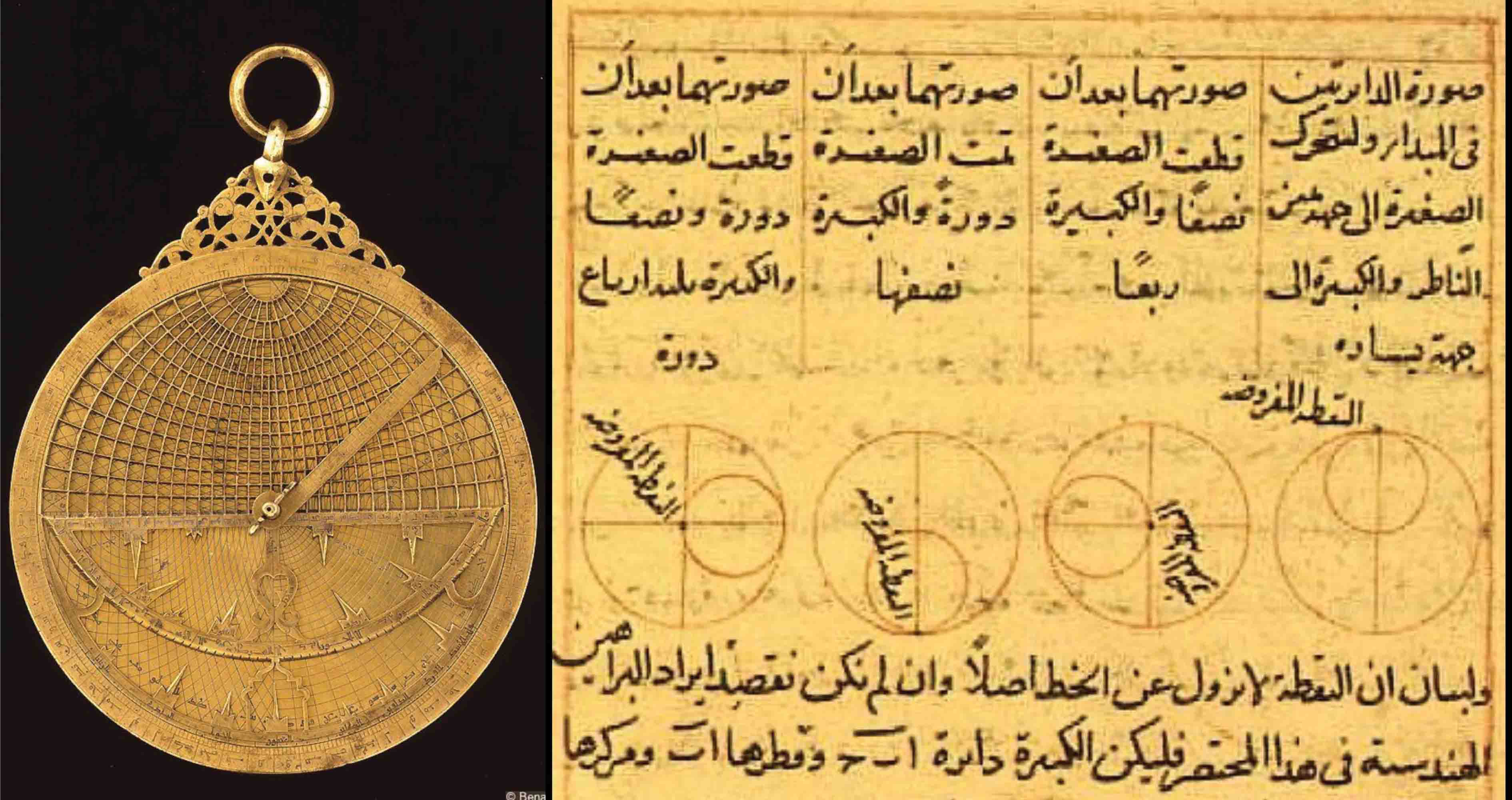}\hfill  
  \caption{Left: “Universal astrolabe,” MWNF database entry (2025) \citep{king1999instruments, morrison2007astrolabe}. Right: Tusi couple, 13th-century sketch attributed to Naṣīr al-Dīn al Tūsī (Vat. Arab. 319) \citep{ragep2001freeing}.}
  \label{fig:inst_astrolabe_tusi}
\end{figure}
Astronomy also served daily religious life through the profession of \textit{muwaqqits} (mosque timekeepers). They designed sundials, compiled and applied \textit{zij} tables, and taught practical astronomy in mosques. The 14th-century \textit{muwaqqit} Ibn al-Shatir at the Umayyad Mosque in Damascus built a marble sundial (1371–72) and constructed planetary models whose mathematics closely parallels Copernican schemes—whether by transmission or independent derivation remains debated \citep{north2008cosmos, ragep2001freeing}.

Mechanical innovation accompanied observation. In 1206, al-Jazari described the water-powered “Castle Clock,” displaying zodiac, solar, and lunar indicators. In the 1570s, Taqi al-Din installed precise mechanical clocks at the Istanbul Observatory for transit timing, a major step for positional accuracy, although the observatory’s 1580 destruction curtailed this line of development \citep{north2008cosmos}.

Arabic-labeled astrolabes, \textit{zij} tables, and practices flowed into Christian Spain, Sicily, and Italy via translation centers such as Toledo; Gerard of Cremona translated al-Zarqali’s tables, while Plato of Tivoli translated al-Battani’s works \citep{burnett2009toledo}. Tusi’s kinematic couple and Ibn al-Shaṭir’s planetary models surface in Copernicus’s \textit{De Revolutionibus}, showing active integration of Islamic techniques into Renaissance astronomy \citep{ragep2001freeing}.

\begin{table}[h]
\centering
\caption{Major astronomical instruments of the Islamic Golden Age.}
\label{tab:IslamicInstruments}
\begin{tabular}{@{}p{3cm}p{2.5cm}p{6.2cm}p{3.3cm}@{}}
\toprule
Instrument & Period / Scale & Primary functions & Example / Source \\
\midrule
Astrolabe & 8th–15th c., portable & Local time, altitude, dawn, qibla, teaching & Al-Fazārī, al-Sūfī; museum survivals \citep{king1999instruments, morrison2007astrolabe} \\
Sine quadrant (\textit{rub al-mujayyab}) & 10th–15th c., handheld & Trigonometric solutions for altitude, prayer times, qibla & Standard mosque and school tool \citep{king1999instruments} \\
Armillary sphere / mural quadrant & 10th–16th c., fixed & Positional astronomy, declinations, right ascensions & Marāgha, Samarkand \citep{north2008cosmos} \\
Fakhri Sextant (Rayy) & 994 CE, monumental & Solstitial altitude, obliquity, latitude & Al-Khujandī; refined by al-Bīrūnī \citep{albiruni1954qanun} \\
Mechanical clocks & 1206–1580, large & Astronomical display, transit timing & Al-Jazarī; Taqī al-Dīn (Istanbul) \citep{north2008cosmos} \\
\bottomrule
\end{tabular}
\end{table}

These instruments show how Islamic astronomy integrated practical needs, religious requirements, and theoretical advances within a single program, and how its technologies and methods entered and shaped medieval and early modern scientific traditions beyond the Islamic world.

\section{Global Comparison: Islamic World, Chinese, and European Astronomy}

To appreciate the Islamic astronomical tradition within the broader scope of world astronomy, it must be viewed as part of an interconnected global dialogue. While Muslim astronomers extended Greek, Persian, and Indian legacies and transmitted new knowledge to Europe, other civilizations such as China and Europe were simultaneously advancing distinctive approaches. The global history of astronomy thus forms a continuous conversation across cultures.

\subsection{China: Long Records and Celestial Bureaucracy}
Chinese astronomy emphasized meticulous long-term recordkeeping of celestial events. Imperial court astronomers maintained continuous observational archives for more than two millennia. Among the most famous is the detailed record of the \textbf{Supernova of 1054 CE}, which created the Crab Nebula. Chinese accounts describe its sudden appearance near Taurus, visible in daylight for 23 days—one of the most precise pre-telescopic stellar records \citep{stephenson2003earthrotation}. These data remain invaluable to modern astrophysics for modeling supernova evolution.

China’s strength lay in institutional continuity. The \textit{Astronomical Bureau} systematically documented observations and integrated them into the imperial calendar. The water-driven clock tower of \textbf{Su Song} (1090 CE) incorporated an escapement and chain drive, uniting celestial display with mechanical ingenuity. Scholars like \textbf{Shen Kuo} (1031–1095) noted magnetic declination, recognizing that the compass needle does not align with true north. Chinese astronomers excelled in eclipse prediction and calendar design, prioritizing administrative precision and practical application \citep{needham1959sciencechina3}.

\subsection{The Islamic World: Instruments, Mathematics, and Transmission}
Islamic astronomy developed a distinctive synthesis of mathematical innovation, instrumentation, and institutional collaboration. Rather than serving as a conduit between Greek and European traditions, it constituted an independent scientific revolution.

From the 8th century onward, Muslim astronomers adapted Greek and Indian sources while introducing original techniques. The \textbf{astrolabe} and \textbf{sine quadrant} became ubiquitous, serving both scientific and devotional purposes such as qibla determination and prayer timing \citep{king1999instruments}. Monumental devices like the \textbf{Fakhri Sextant} (Rayy, 994 CE) and Ulugh Beg’s 15th-century sextant at Samarkand (radius $\sim$40 m) achieved extraordinary pre-telescopic accuracy \citep{albiruni1954qanun}.

Institutions such as the \textbf{Marāgha Observatory} (1259) under Nasīr al-Dīn al-Tūsī pioneered collaborative research, housing extensive libraries and introducing mathematical devices like the Tūsī couple (see Fig.~\ref{fig:inst_astrolabe_tusi}), later echoed in Copernicus’s \textit{De Revolutionibus} \citep{ragep2001freeing}. Mosque astronomers (\textit{muwaqqits}) such as \textbf{Ibn al-Shāṭir} refined sundials and planetary models whose geometry anticipated heliocentric formulations. 

This synthesis joined practical religious needs, institutional science, and theoretical reform. By the late 16th century, \textbf{Taqī al-Dīn’s Istanbul Observatory} (1577–1580) advanced precision measurement with mechanical clocks resolving seconds and employing decimal fractions. His methods paralleled those of Tycho Brahe, though his observatory’s destruction ended the Ottoman program prematurely. Through translation centers at Toledo and Sicily, works by al-Battānī, al-Zarqālī, and Ibn al-Shāṭir shaped Renaissance astronomy \citep{burnett2009toledo}.

\subsection{Europe: From Monastic Schools to Scientific Revolution}
European astronomy evolved in different institutional settings, first in monasteries and cathedral schools, later in universities. Translations of Arabic texts in the 12th and 13th centuries provided the mathematical and observational foundations for systematic study. The \textbf{Alfonsine Tables} (1252) of Toledo, compiled under Alfonso X, exemplify this synthesis.

By the 16th century, European astronomers combined inherited mathematical models with new instruments and institutions. \textbf{Tycho Brahe’s} observatory achieved arcminute precision using giant mural quadrants; the \textbf{Kassel program} under Wilhelm IV pursued a highly accurate stellar catalog. \textbf{Copernicus} employed kinematic mechanisms like the Tusi couple in proposing heliocentrism, and \textbf{Kepler} used Tycho’s data to derive elliptical orbits. The invention of the telescope in 1609 marked Europe’s transformation of inherited frameworks into a new cosmology.

\subsection{Bridging Civilizations: Cross-Cultural Exchanges}
Far from isolated, these traditions were interlinked through travel, trade, and diplomacy. During the Mongol era, Islamic astronomy entered China: in 1267, the Persian scholar \textbf{Jamāl al-Dīn} presented instruments at Kublai Khan’s court, leading to the \textit{Huihui Lifa} (Islamic Astronomical Tables) that merged Ptolemaic and Chinese methods. This collaboration produced Yuan dynasty calendars that integrated Islamic mathematical precision \citep{sayili1960observatory}.

Conversely, Arabic texts translated in Toledo and Sicily catalyzed Europe’s Renaissance, while Jesuit missions in the 17th century reintroduced telescopic astronomy to China, reforming imperial calendars. Ottoman and Safavid scholars engaged with both Islamic and European techniques, maintaining active observatories and translation programs. 

These multidirectional exchanges underscore astronomy’s global evolution as a cooperative enterprise. Chinese recordkeeping ensured observational continuity; Islamic astronomy contributed mathematical rigor and instrumentation; and European astronomy restructured inherited ideas into the modern scientific paradigm. Together, these civilizations built the cumulative foundation of modern astrophysics through shared curiosity and continuous cross-cultural transmission.

\section{Transformation in Islamic Astronomy}

Islamic astronomy emerged in the 8th century and continued to evolve through the 16th century and beyond. Its development was dynamic, shaped by political, economic, and intellectual transformations. The tradition adapted to new institutional and ideological settings while sustaining innovation. Mathematical devices by Nasīr al-Dīn al-Tūsī that appear in Copernican kinematics likely diffused indirectly, while Ibn al-Shāṭir’s planetary models closely parallel Copernican systems \citep{ragep2001freeing}. Taqī al-Dīn’s late sixteenth-century Istanbul Observatory deployed precision instruments and a second-resolving mechanical clock rivaling European advances, although its short operation limited long-term data. Rather than viewing this history as one of linear rise and fall, it is better understood as a process of transformation—scientific practice migrating and adapting across cultural and temporal boundaries \citep{saliba2007islamicScience, dallal2010islamScience}.

\subsection{Methodological and Intellectual Evolution}

In the early Golden Age, Muslim scholars translated and expanded upon Greek, Indian, and Persian sources. By the 11th century, debates intensified around methodology. The theologian \textbf{al-Ghazālī} (1058–1111), in \textit{The Incoherence of the Philosophers} (\textit{Tahāfut al-Falāsifa}), challenged Avicennian metaphysics and questioned causal necessity in natural phenomena \citep{alghazali2000tahafut}. His critique fostered occasionalism—the view that divine will directly governs all events. While this framework limited speculative cosmology, it did not obstruct empirical astronomy. Scholars continued refining calendars, qibla determination, and timekeeping, maintaining a rigorous observational tradition. 

Comparison with Europe reveals different trajectories: while Islamic scholars debated metaphysical causation, European universities increasingly formalized natural philosophy, creating conditions for later institutionalized science. Both traditions negotiated religious and philosophical boundaries while expanding empirical inquiry.

\subsection{Institutional Evolution and Adaptation}

Political change reshaped but did not extinguish astronomical practice. The fragmentation of Abbasid authority after the 10th century curtailed unified patronage, but new centers arose under diverse rulers. The Mongols established the \textbf{Marāgha Observatory} (1259), demonstrating institutional renewal after the fall of Baghdad. Later, \textbf{Ulugh Beg} founded the \textbf{Samarkand Observatory} (1420s), producing stellar catalogues of exceptional precision—errors of only a few arcminutes—remarkable achievements for pre-telescopic astronomy \citep{sayili1960observatory}. Despite Ulugh Beg’s assassination in 1449, his data influenced later developments.

The Ottoman Empire institutionalized astronomy through the post of Chief Astronomer and networks of \textit{muwaqqithānes} (timekeeper offices). Taqī al-Dīn’s Istanbul Observatory (1577–1580) embodied this maturity, equipped with advanced instruments and mechanical clocks matching European precision \citep{north2008cosmos}. In Europe, by contrast, universities provided enduring institutional continuity. Both civilizations adapted to political realities through distinct but equally effective models.

\subsection{Economic Adaptation and Resource Allocation}

Shifts in global trade transformed scientific priorities. European powers, driven by maritime expansion, invested heavily in navigational astronomy—culminating in the \textbf{Greenwich Observatory} (1675)—while Islamic empires like the Ottomans integrated astronomy into administration, religion, and military logistics. Patronage patterns reflected economic structures: European states funded oceanic exploration, whereas Islamic states supported land-based governance and calendar regulation. Both models demonstrate rational adaptation to distinct strategic needs rather than intrinsic superiority.

\subsection{Technological Integration and Innovation Paths}

Technological innovation followed different institutional logics. Telescopic astronomy, emerging in Europe after 1609, appeared sporadically in Islamic scholarship but was not systematically institutionalized. Printing faced distinct challenges in Arabic script, delaying widespread adoption. When \textbf{Ibrahim Müteferriqa} founded an Ottoman press (1727–1729), he built upon centuries of manuscript sophistication \citep{eisenstein1980printing}. Meanwhile, Taqī al-Dīn’s decimal mechanical clocks represented technological ambition equal to Europe’s, illustrating continued inventive capacity within adapted institutional frameworks.

\subsection{Knowledge Networks and Scientific Transformation}

The parallels between Copernican models and Islamic precursors underscore the permeability of scientific exchange. The \textbf{Tusi couple} and \textbf{Ibn al-Shāṭir’s} models informed European reformulations, though exact transmission routes remain debated \citep{burnett2009toledo}. Simultaneously, Islamic scholars advanced instrumentation and computation. These developments occurred within emerging global knowledge networks that linked scholars across Eurasia. 

\begin{figure}[h]
    \centering
    \includegraphics[width=0.8\linewidth]{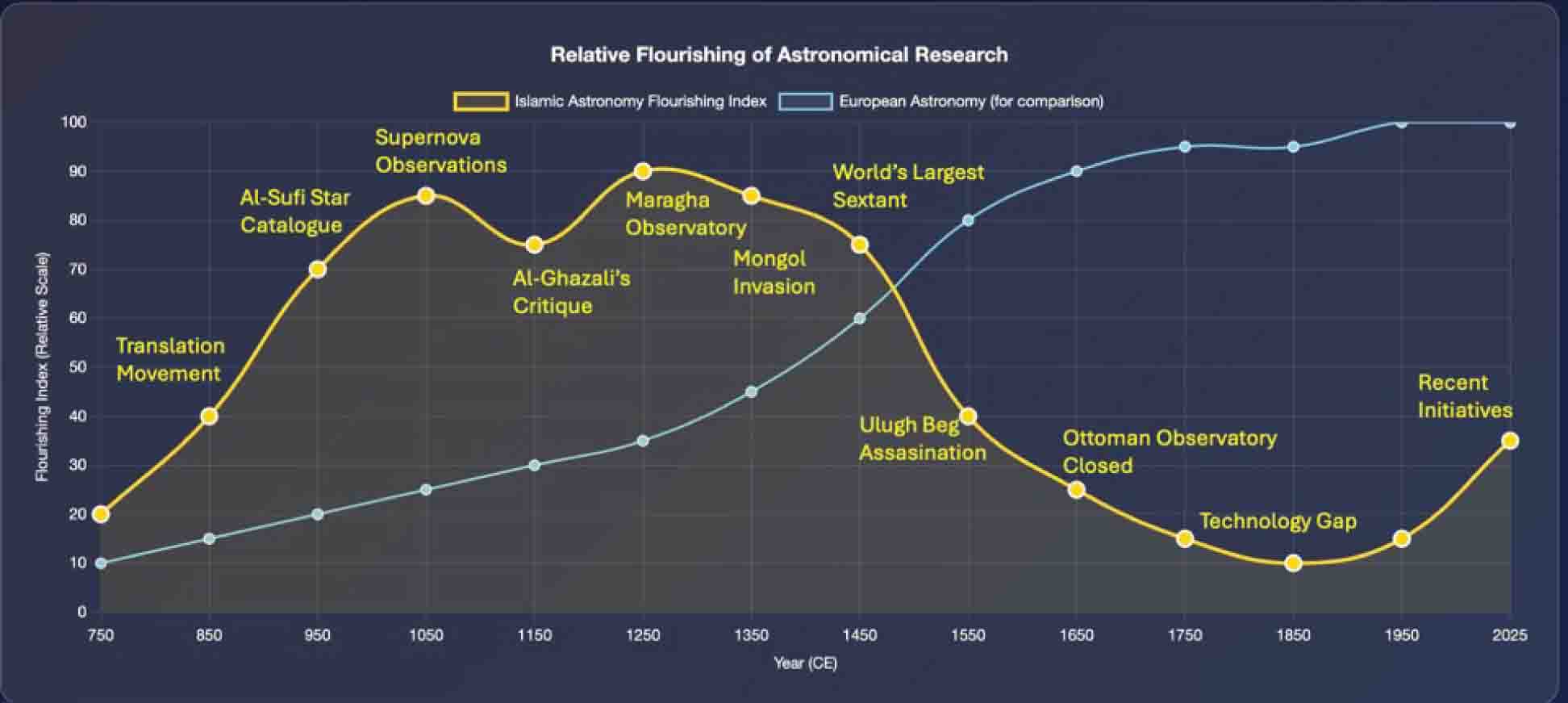}
    \caption{Transformation and continuity of Islamic astronomy (750–2025~CE). The schematic highlights major institutional, intellectual, and technological phases: classical development (8th–14th c.), transformation (15th–17th c.), and modern revival (20th–21st c.). The curve represents a qualitative and conceptual “flourishing index,” synthesizing evidence from institutional activity, discoveries, and global integration. The chart emphasizes that scientific evolution is cyclical and context-dependent, not linearly progressive. The index is created by the authors as a qualitative demonstration.}
    \label{fig:islamic_transformation}
\end{figure}

Islamic astronomy’s transformation reflects not decline but adaptation—its practitioners continuously renegotiated political patronage, intellectual paradigms, and technological tools. Figure~\ref{fig:islamic_transformation} illustrates the continuity and transformation of Islamic astronomy from 750 to 2025~CE, emphasizing cyclical phases of innovation, adaptation, and renewal across more than a millennium.

\section{Contemporary Revival: Astronomy and Space in the Islamic World}
The legacy of Islamic astronomy continues in modern scientific and institutional initiatives. Across North Africa, the Middle East, and Central Asia, renewed astronomical activity fuses heritage with innovation, balancing national ambition and global cooperation. This revival reflects the spirit of medieval centers such as Baghdad, Marāgha, and Samarkand, but faces new challenges: uneven funding, geopolitical tensions, and the need for sustainable institutions.

\subsection{Technological Innovation and Religious Practice}

\begin{figure}[h]
  \centering
  \includegraphics[width=.6\linewidth]{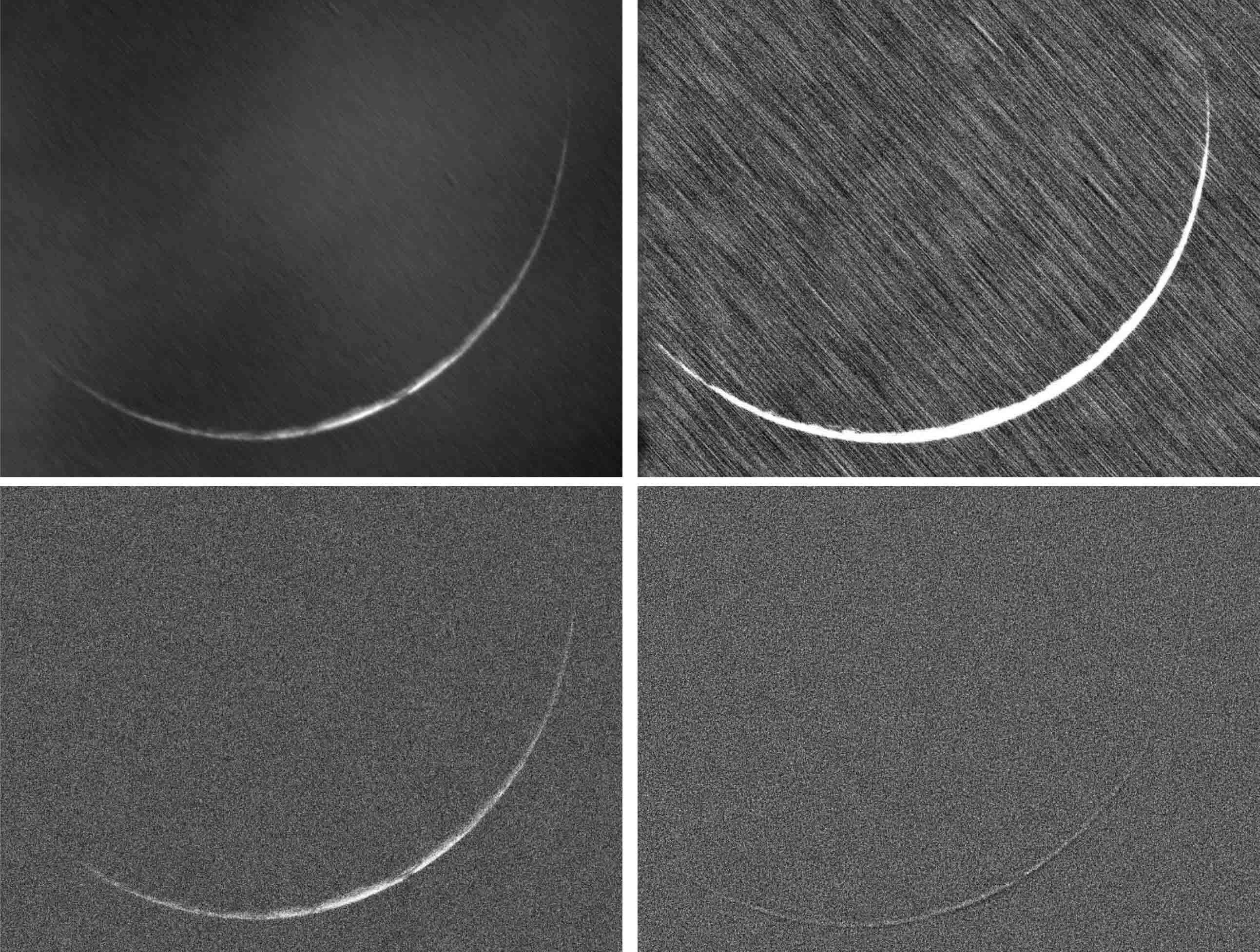}
  \caption{Daylight lunar crescent observation techniques developed in the Islamic world for calendar determination. (Images from the Mecca Clock Tower Museum).}
  \label{fig:lunar_crescent}
\end{figure}

Modern Islamic astronomy continues to unite scientific precision with religious needs. Saudi Arabia’s \textbf{King Abdulaziz University} operates the \textit{Institute for Astronomy and Geophysics}, conducting advanced lunar visibility research to determine the Islamic calendar. At the \textbf{Mecca Clock Tower Observatory}, researchers designed a modified telescope with a long optical tube and solar filters to detect the thin crescent Moon in daylight, allowing unprecedented accuracy in determining Ramadan, Hajj, and Eid. Figure~\ref{fig:lunar_crescent} shows examples of these observations.
The \textbf{International Astronomical Center} in Abu Dhabi standardizes the daylight crescent observation protocols using specialized solar filters, advancing lunar astronomy and calendar science \citep{itu2023program}. These initiatives show how centuries-old Islamic traditions continue to inspire technological innovation, integrating spiritual observance with cutting-edge optics and data processing.
\subsection{International Collaboration and Regional Contributions}

Observatories around the Islamic world now participate in global research networks. Morocco’s \textbf{Oukaïmeden Observatory} hosts the TRAPPIST-North telescope, contributing to exoplanet studies and the ATLAS asteroid survey network \citep{nasaesa2024partnerships}. Regional educational success is evident in the multiple medals earned by students from Iran, Turkey, and other nations at the International Olympiad on Astronomy and Astrophysics \citep{iac2024lunar}.

The \textbf{UAE’s Emirates Mars Mission (Hope)}, operational since 2021, continues to provide vital data on Martian atmospheric dynamics and auroras \citep{sharaf2021hope}. These projects illustrate how Islamic world institutions are reintegrating into planetary and astrophysical research, their long-term success hinging on consistent institutional support.

\subsection{Educational Development and Infrastructure}

Universities such as \textbf{King Abdulaziz University} and the \textbf{American University of Sharjah} have expanded astronomy programs connecting historical Islamic astronomy to modern astrophysics \citep{spa2023ssaTransform}. Public outreach through Olympiad training, stargazing events, and planetaria increases scientific literacy.

Challenges remain: limited access in rural regions, funding shortfalls, and insufficient integration of artificial intelligence, computational modeling, and space systems. Addressing these infrastructural issues is essential for sustaining scientific growth.

\subsection{Space Program Development}

Islamic nations are emerging participants in global space science. The \textbf{UAE} shares Mars mission data through international repositories. \textbf{Saudi Arabia} founded the \textbf{Saudi Space Agency} in 2023, joined the Artemis Accords in 2022, and partnered in ISS missions. \textbf{Turkey} launched its first astronaut mission in 2024 and announced lunar exploration goals. \textbf{Iran} continues satellite and remote-sensing development despite sanctions.

These programs link regional agencies to global observatories and missions, including data collaboration with the James Webb Space Telescope and asteroid monitoring networks, signifying a transition from knowledge reception to active contribution.

The future of astronomy and space in the Islamic world depends on sustained investment and robust institutional frameworks. Long-term success requires continuous funding, technical expertise, and integration into global collaborations. Key challenges include maintaining political continuity, developing domestic manufacturing, and building reliable scientific careers.

\subsubsection*{Case Study: Saudi Arabia’s Transformation}

Saudi Arabia exemplifies the intersection of tradition, modernization, and ambition. From early lunar observation programs to contemporary space initiatives, its trajectory shows increasing sophistication. King Faisal’s establishment of observatories in the 20th century laid the groundwork for today’s expansion under \textbf{Vision 2030}. The \textbf{Saudi Space Agency}, upgraded in 2023, integrates national goals in astronomy, planetary science, and aerospace engineering.

The \textbf{Public Investment Fund (PIF)} launched the \textbf{Neo Space Group (NSG)} in 2024, focusing on satellite communications, navigation, and venture capital for aerospace innovation. Partnerships with NASA have yielded historic milestones, including the 2023 Axiom-2 mission that sent Rayyanah Barnawi—the first Saudi woman in space—and Ali AlQarni to the ISS. Subsequent cooperation agreements and participation in Artemis mark Saudi Arabia’s transition from lunar calendar studies to space exploration \citep{nasaesa2024partnerships}.

Manufacturing partnerships, such as those with Halo Space for near-space tourism, demonstrate ambitions extending beyond exploration to aerospace production and commercial applications. Saudi Arabia’s clear skies and strategic geography strengthen its role as a hub for observation and launch infrastructure.

The revival of astronomy and space science in the Islamic world represents continuity with its intellectual heritage and a decisive step toward global scientific participation. Whether these developments become enduring achievements or transient prestige projects will depend on institutional resilience, educational reform, and sustained international collaboration.

\section{Conclusion: From the ``Little Cloud'' to New Horizons}

More than a thousand years ago, \textbf{Abd al-Rahman al-Sufi} gazed at a faint smudge in Andromeda and called it a “little cloud.” With nothing more than the naked eye, he captured what we now know to be an entire galaxy about 2.5 million light-years away. That simple observation, preserved in the \textit{Book of Fixed Stars} (964 CE), embodies the empirical precision and methodological rigor that characterized Islamic-era astronomy at its height—a tradition that transformed inherited knowledge into systematic science through careful observation, mathematical innovation, and instrumental development.

From the 8th through the 16th centuries, scholars such as \textbf{al-Biruni}, \textbf{al-Battani}, \textbf{Ibn Sina}, \textbf{Nasir al-Din al-Tusi}, and \textbf{Taqi al-Din} combined mathematical precision, systematic observation, and technological innovation to advance astronomical understanding \citep{albiruni1954qanun, dhanani1994kalam, ragep2001freeing}. Their \textit{zijes}, astrolabes, mechanical clocks, and monumental sextants not only served practical needs—such as qibla determination, calendar calculation, and prayer timing—but also created methodological foundations that became essential to global astronomical development. Islamic-era mathematical models are evident in Copernican kinematics and geometry, with routes of transmission still debated \citep{saliba2007islamicScience, north2008cosmos}. Their observational precision influenced Renaissance astronomy, and their institutional models informed early modern scientific practice.

This tradition underwent a complex transformation rather than a simple decline. Political changes reshaped patronage structures; economic shifts altered resource priorities; and intellectual developments refined methodological approaches while maintaining theoretical innovation \citep{dallal2010islamScience, huff2017curiosity}. The westward transmission of Islamic astronomical knowledge through translation centers represented a major achievement of global knowledge networks rather than knowledge loss \citep{burnett2009toledo}. Even as European institutions developed telescopic astronomy and printing technology, Islamic scholars such as \textbf{Taqi al-Din} continued advancing precision instrumentation and mathematical techniques that matched contemporary European achievements in scope and ambition \citep{sayili1960observatory}.

Contemporary revival efforts across the Islamic world demonstrate renewed institutional capacity and international engagement. Morocco’s \textbf{Oukaimeden Observatory} contributes meaningfully to global exoplanet research and asteroid tracking networks \citep{benkhaldoun2021oukContrib}. The UAE’s \textbf{Hope Mars Mission} provides valuable atmospheric data while demonstrating sophisticated space science capabilities \citep{alamiri2022emm}. Educational achievements through international astronomy olympiads reflect strong foundational preparation \citep{ioaaResults}, while expanding university programs—from Istanbul to Sharjah—create new generations of astronomers. Saudi Arabia’s \textbf{Space Agency}, astronaut missions, and international partnerships signal serious commitment to space science development \citep{ssa2024strategy}.

Yet these developments face substantial challenges that historical analysis helps illuminate. Funding remains concentrated in showcase projects rather than distributed institutional networks. Geopolitical tensions restrict international collaboration essential for competitive research. Sanctions, visa restrictions, and technology transfer limitations create barriers unknown to medieval scholars. Most critically, institutional sustainability depends heavily on political patronage—the same vulnerability that limited the long-term impact of medieval observatories \citep{oecd2023sti, brookings2024meSpace}.

For contemporary revival efforts to achieve lasting scientific significance, three strategic priorities emerge from this historical analysis:

\paragraph{1. Establish shared digital infrastructure and grassroots networks.}
Develop virtual observatory consortiums enabling real-time data sharing between regional institutions. Support astronomy club networks and community formation across borders. Create online educational platforms offering advanced astronomical training, and establish computational cloud resources for collaborative research projects. Such initiatives would rebuild distributed scientific networks that mirror the cross-border collaborations sustaining medieval Islamic astronomy.

\paragraph{2. Pursue targeted participation in major international projects.}
Rather than building comprehensive national programs, focus on contributing specialized expertise to large-scale efforts such as the Square Kilometre Array (SKA), Extremely Large Telescope (ELT) initiatives, and planetary exploration missions. Contributions may include atmospheric modeling, precision optics, or mathematical analysis algorithms—strategies that establish measurable research visibility while managing costs effectively.

\paragraph{3. Integrate astronomical education with national technology priorities.}
Create structured career pipelines linking astronomical data analysis with artificial intelligence, satellite engineering with space technology, and climate modeling with computational methods. Develop dual-degree programs that ensure graduates contribute both to regional economic development and scientific advancement.

The historical trajectory from medieval Islamic achievements through institutional transformation to modern revival reveals fundamental patterns of scientific development under challenging conditions. While contemporary revival efforts show genuine promise, they face geopolitical realities absent from the medieval period. Success requires acknowledging these constraints while designing resilient institutional solutions.

Most significantly, this analysis reveals that contemporary Islamic astronomy’s path forward lies not in reclaiming historical dominance, but in cultivating sustainable contributions to humanity’s cosmic understanding within current global frameworks. Historical Islamic astronomy succeeded by adapting inherited knowledge to serve both practical needs and intellectual curiosity—approaches that remain vital for navigating modern challenges.

From al-Sufi’s naked-eye observation of the distant “little cloud” to the data collected by space-based observatories, the continuity of human curiosity about the universe provides both inspiration and guidance. The contributions of the Islamic world demonstrate that scientific achievement arises from cultural, political, and economic systems that foster systematic inquiry and collaboration. Whether contemporary institutions can adapt these lessons will determine whether this revival represents a genuine renewal of the legacy of Islamic astronomy in humanity’s shared quest to understand the cosmos.

\section*{Acknowledgments}
Authors would like to thank Farshid Maghami Asl and Kian Maghami Asl for their contributions to this research. We also thank the time of the Maghami Asl family’s private observatory which was provided to the authors for this research. 

\bibliographystyle{unsrtnat}
\bibliography{references}

\end{document}